\documentclass[manuscript,flushrt,showpacs,preprintnumbers,amsmath,amssymb]{revtex4}
\usepackage{graphicx}
\usepackage{latexsym}
\usepackage{color}
\usepackage{dcolumn}
\usepackage{bm}
\usepackage{natbib}
\usepackage{setspace}

\begin{document}

\title{The Origin of Magnetic Fields in Galaxies}

\author{Rafael S. de Souza}
\email{Rafael@astro.iag.usp.br}
\author{Reuven Opher}%
 \email{Opher@astro.iag.usp.br}
\affiliation{IAG, Universidade de S\~{a}o Paulo, Rua do Mat\~{a}o
1226, Cidade Universit\'{a}ria, CEP 05508-900, S\~{a}o Paulo, SP,
Brazil.}

\begin{abstract}

Microgauss magnetic fields are observed in all galaxies at low and high redshifts. The origin of these intense magnetic fields is a challenging question in astrophysics.  We show here that the natural plasma fluctuations in the primordial universe (assumed to be random), predicted by the Fluctuation-Dissipation-Theorem, predicts $\sim 0.034~ \mu G$ fields over $\sim 0.3$ kpc regions in galaxies. 
 If  the dipole magnetic fields predicted by the Fluctuation-Dissipation-Theorem are  not completely random,  microgauss fields over regions $\gtrsim 0.34$ kpc are easily obtained. The model is thus a strong candidate for resolving  the problem of the origin of magnetic fields  in $\lesssim 10^{9}$ years  in high redshift galaxies.
 \end{abstract}

 \pacs{98.54.Kt, 98.62.En} 

\maketitle
\section{Introduction}

The origin of large-scale cosmic magnetic fields in galaxies and protogalaxies remains   a challenging problem  in astrophysics \citep{zwe97, kul08, raf08, wid02}. There have been many attempts to explain  the origin of cosmic magnetic  fields.
One of the  first popular  astrophysical theories to create seed fields was   the Biermann
mechanism \citep{bie50}. It has been suggested that this mechanism
acts in diverse astrophysical systems, such as  large scale
structure formation \citep{pee67,ree72,was78}, cosmological ionizing
fronts \citep{gne00},   star formation and  supernova explosions  \citep{mir98, han05}.  \citet{ryu08} made simulations showing that cosmological shocks can create   average magnetic fields of a few $\mu G$ inside cluster/groups,  $ \sim 0.1 ~\mu G$ around clusters/groups, and $\sim 10~nG$ in filaments.  \citet{med06} showed that magnetic fields can be produced by collisionless shocks in galaxy clusters and in the intercluster medium (ICM) during large scale structure formation. \citet{ars09} studied the evolution of magnetic fields  in 
galaxies coupled with  hierarchical structure formation.   \citet{ich06}   investigated  second-order couplings between photons and electrons  as a possible origin of  magnetic fields on cosmological scales before the epoch of recombination.  The creation of early magnetic fields generated by cosmological perturbations have also been investigated  \citep{tak05, tak06, cla01, mae09}.


In our galaxy, the magnetic field is coherent over kpc scales with alternating directions in the arm and inter-arm regions (e.g.,\citet{kro94, han08}). Such alternations are expected for magnetic fields of primordial origin \citep{gra01}.

Various observations put upper limits on  the intensity of  a homogeneous primordial magnetic field. Observations of the small-scale cosmic microwave background (CMB) anisotropy yield an upper comoving limit of $4.7 ~nG$ for a homogeneous primordial field \citep{yam06}. Reionization of the Universe puts upper limits of $0.7-3~ nG$ for a homogeneous primordial field, depending on the assumptions of the stellar population that is responsible for reionizing the Universe \citep{sch08}. Another upper limit for a homogenous primordial magnetic field is the magnetic Jeans mass $\sim 10^{10} M_{\odot} ~(B/3nG)^{3}$ \citep{sub98, set05}. Thus, if we are investigating the collapse of a $\sim 10^{7} M_{\odot}$ protogalaxy, the homogeneous primordial magnetic field must be $< 0.3~ nG$ in order for collapse to occur.  

Galactic magnetic fields have been  suggested to  have  evolved in three main stages.  In the first stage,   seed fields were embedded in the protogalaxy.   They may have had   a   primordial origin,  as suggested in this paper. Another possibility is that the seed fields could have been injected  into the protogalaxies by AGN jets, radio lobes, supernovas,  or a combination of the above.  Still another possibility is that  the seed fields may  have been   created by the Biermann battery  during the formation of the protogalaxy.  In the second   stage, the seed  fields were amplified by compression,   shearing flows, turbulent flows, magneto-rotational instabilities,  dynamos or by a combination of the above. In the last stage magnetic fields were ordered by a large scale dynamo \citep{beck06}.


\citet{ryu08} investigated the amplification of magnetic fields due to turbulent vorticity created at cosmological shocks during the formation of large scale structures.  A given vorticity $\omega$ can be characterized by a characteristic velocity $V_{c}$ over a characteristic distance $L_{c}$. \citeauthor{ryu08} found  that $\omega$ typically is
\begin{equation}
\omega \sim 1-3 \times 10^{-16} s^{-1},
\end{equation}
which corresponds to 10-30 turnovers in the age of the universe. They investigated $L_{c} > $ 1 Mpc $h^{-1}$. We investigate $L_{c} \simeq$ 200 kpc $h^{-1}$ in  protogalaxies for  a  similar vorticity.

We show that a seed field $0.003 ~nG$ over a comoving $2~ kpc$ region at  \emph{z} $\sim 10$,  predicted by the Fluctuation-Dissipation Theorem \citep{raf08},  amplified by the small scale  dynamo is a good candidate for the origin of magnetic fields in galaxies. \citet{sub97, sub99} and \citet{bra00} derived the non-linear
evolution equations for the magnetic correlations.   We use their formulation for the small scale  dynamo  and solve the nonlinear  equations numerically.  In \S II,  we review  the creation of  magnetic fields due to electromagnetic fluctuations in hot dense equilibrium primordial plasmas,  as described in our previous work \citep{raf08}.  In \S III, we discuss the   small scale  dynamo and in \S IV,   the important parameters of the plasma to be used in the calculations.  In \S V,   we present our results and in \S VI our conclusions.

\section{Creation of Magnetic Fields Due to  Electromagnetic Fluctuations in Hot Dense  Primordial Plasmas in Equilibrium}

Thermal electromagnetic fluctuations are present in all plasmas,
including those in thermal equilibrium. The level of  the fluctuations  is
related to the dissipative characteristics of the plasma, as
described by the Fluctuation-Dissipation Theorem (FDT) \citep{kub57}
[see also \citet{akh75,sit67,Ros65,daw68}].

\citet{raf08} studied the evolution of these bubbles as the Universe expanded and  found that the magnetic fields in the  bubbles, created
originally at the quark-hadron phase transition (QHPT), had a value $\sim 9 ~ \mu$G  and a size 0.1 pc at the redshift
$\emph{z} \sim 10$ (see Table 1 of \citep{raf08}).    Assuming that  the fields are randomly oriented, the average magnetic field over a  region D is $B = 9 \mu G ~(0.1pc/D)^{3/2}$.  The theory thus predicts an average magnetic field $0.003 ~nG$ over a $2 ~kpc$ region at $\emph{z} \sim 10$. We assume this seed field and examine its amplification in a 
 protogalaxy by the small scale  dynamo, discussed in the next section.

\section{Small Scale  Dynamo}

In a partially ionized medium,  the magnetic field evolution is
governed by the induction equation
\begin{equation}
(\partial {\bf B}/ \partial t) =
{\bf \nabla } \times ( {\bf v}_i \times {\bf B} -
 \eta {\bf \nabla } \times {\bf B}),
\end{equation}
where ${\bf B}$ is the magnetic field,
${\bf v}_i$ the velocity of the ionic component of the fluid and
$\eta$ is  the ohmic resistivity.

Let   $L_c$ be the coherence scale
of the turbulence. Consider a system whose size is $> L_{c}$ where  the mean field, averaged over any scale, is
negligible. We take $\mathbf{B}$ to be a homogeneous, isotropic, Gaussian
random field with a negligible mean average value. For equal time, the two
point correlation of the magnetic field is
\begin{equation}
\left\langle B^{\,i}\left( \mathbf{x},t\right) B^{\,j}\left( \mathbf{y}
,t\right) \right\rangle =M^{\,ij}\left( r,t\right),
\end{equation}\label{c}
where
\begin{equation}
M^{\,ij}=M_{N}\left[\delta ^{\,ij}-\left( \frac{r^{\,i}r^{\,j}}{r^{\,2}}\right)
\right] +M_{L}\left( \frac{r^{\,i}r^{\,j}}{r^{\,2}}\right) +H\epsilon _{\,ijk}\,r^{\,k},
\label{d}
\end{equation}
\citep{sub97, sub99, bra00}. 
$M_{L}\left( r,t\right) $ and $M_{N}\left( r,t\right) $ are the longitudinal
and transverse correlation functions, respectively, of the magnetic field and
$H\left(r,t\right) $ is the helical term of the correlations.  Since $
\mathbf{\nabla \cdot{B}}=0,$ we have $M_{N}=\left( 1/2\,r\right) \partial
\left( r^{\,2}M_{L}\right) /\left( \partial r\right) $ \citep{mon75}. The
induction equation can be converted into evolution equations for $M_{L}$ and
$H:$
\begin{eqnarray}
\frac{\partial M_{L}}{\partial \,t}\left( r,t\right) &=&\frac{2}{r^{\,4}}\frac{
\partial }{\partial
\,r}\left( r^{4}\kappa _{N}\left( r,t\right) \frac{
\partial M_{L}\left( r,t\right) }{\partial \,r}\right)  \nonumber \\
&+& G(r)M_{L}\left( r,t\right) +4\,\alpha _{N}H\left( r,t\right),  \label{ML}
\end{eqnarray}
and
\begin{eqnarray}
\frac{\partial H}{\partial {\,t}}\left( r,t\right) &=& \frac{1}{r^{\,4}}\frac{
\partial }{\partial {\,r}}\left[ r^{\,4}\frac{\partial }{\partial {\,r}}\left[ \,2\,\kappa
_{N}\left( r,t\right) H\left( r,t\right) \right. \right. \nonumber\\
&-& \left. \left. \alpha _{N}\left( r,t\right) M_{L}\left( r,t\right)\right]\, \right] ,
\label{f}
\end{eqnarray}
where
\begin{equation}
\kappa _{N}\left( r,t\right) =\eta +T_{LL}\left( 0\right) -T_{LL}\left(
r\right) +2\,a\,M_{L}\left( 0,t\right)  \label{g},
\end{equation}
\begin{equation}
\alpha _{N}\left( r,t\right) =2\,C\left( 0\right) -2\,C\left( r\right)
-4\,a\,H\left( 0,t\right)  \label{h},
\end{equation}
and
\begin{equation}
G\left( r\right) =-4\left\{ \frac{d}{d\,r}\left[ \frac{T_{NN}\left( r\right) }{
r}\right] +\frac{1}{r^{\,2}}\frac{d}{d\,r}\left[\, r\,T_{LL}\left( r\right) \right]
\right\}  \label{i}
\end{equation}
\citep{sub97,sub99}. $T_{LL}(r)$ and $T_{NN}(r)$ are the
longitudinal and transverse correlation functions for the velocity
field.
The functions $T_{NN}$ and $T_{LL}$  are then related  in the way described
by \citet{sub99}, which we assume here.  
 These equations  for $M_{L}$ and $H,$ describing the evolution of
magnetic correlations at small and large scales. The effective diffusion
coefficient $\kappa _{N}$ includes microscopic diffusion $(\eta ),$ a scale-dependent
turbulent diffusion $\left[T_{LL}\left( 0\right) -T_{LL}\left( r\right)\right], $
and a ambipolar drift $2aM_{L}\left( 0,t\right) ,$ which is
proportional to the energy density of the fluctuating fields. Similarly, $
\alpha _{N}$ is a scale-dependent $\alpha $ effect, proportional to
$[\,2\,C\left( 0\right)
-2\,C\left( r\right )].$ The nonlinear decrement of the $\alpha $ effect due to ambipolar
drift is  $4aH\left(0,t\right), $ proportional to the mean helicity of the magnetic
fluctuations. The $G\left( r\right) $ term in equation (\ref{ML}) allows for rapid
generation of small scale magnetic fluctuations due to velocity shear
\citep{zel83,kaz68,bra00,sub99}.

This turbulent spectrum simulates Kolmogorov turbulence \citep{vai82}.    
As in the galactic interstellar medium, the protogalactic plasma is expected to have  Kolmogorov-turbulence, driven by the shock waves originating from the instabilities,  associated with gravitational collapse.   


In the galactic context,  we can neglect the coupling term $\alpha_{N}H$ as  a very good  approximation since it is very small and consider only the evolution of $M_{L}$ \citep{sub99}.

For turbulent motions on a  scale $L$ and  a velocity scale $v$, the magnetic Reynolds
number (MRN) is $R_m = vL/\eta$. There is a critical MRN, $R_c \approx 60$,
so that for $R_m > R_c$ \citep{sub99},    modes of the small scale dynamo can be excited. The  fluctuating field, correlated on a
scale $L$, grows exponentially   with a
growth rate $\Gamma_L \sim v/L$ \citep{sub99}.

\section{The Parameters of the Turbulent Plasma}

We use the fiducial parameters,  suggested in the  literature for  the  plasma that was present in the protogalaxy \citep{mal02,sch02}:  total mass M  $\sim 10^{12} ~M_{\odot}$,  temperature T  $\sim 10^{6}$ K, and  size $L_{c} \sim 200 ~kpc$.  The ion kinematic viscosity is $\sim 5\times10^{26}~ cm^{2}/s$,  the Spitzer resistivity   $\eta_{s} = 6.53 \times 10^{12} ~T^{-3/2} \ln\Lambda ~cm^{2}s^{-1}$ $\sim 8 \times 10^{4} ~cm^{2}s^{-1}$, and the typical  eddy velocity $V_{c}\sim 10^{7}~cm/s$.

\section{Results}

In Fig. 1, we evaluate $M_{L}$ for various values of r and in Fig. 2 for various values of $V_{c}$, solving numerically equation (\ref{ML}). 
 In Fig 3 we evaluate the mean value of the magnetic field as a  function of r and t.  Our previous work \citep{raf08} showed that the natural fluctuations of the primordial plasma predicted by the Fluctuation-Dissipation Theorem produces a cosmic web of randomly oriented dipole magnetic fields. The average field over a region $\sim 2$ kpc is predicted to be $0.003 ~nG$. We assume this seed field and examine its amplification by the small scale  dynamo in s protogalaxy. This seed field  corresponds to an $M_{L} (\sim B^{2}) \simeq 10^{-23}~G^{2}$.  Of particular interest is thus
the growth of $M_{L}$ with an  initial value $M_{L0} \sim 10^{-23}~ G^{2}$ in Figs. 1 and 2, for initial magnetic fields $B_{0}\sim3\times10^{-12}(2 kpc/r)^{3/2}~G$  of size $r$ in  Fig. 3.

\section{Conclusions and Discussion}

It was shown previously that the magnetic fields,   created immediately after
the quark-hadron transition, produce  relatively intense magnetic dipole fields on small scales at $\emph{z} \sim 10$ \citep{raf08}. We show here that   the predicted seed fields of size $\sim 2$ kpc and intensity $0.003~ nG$ at \emph{z} $\sim 10$    can be amplified by a small scale  dynamo  in protogalaxies to intensities  close to observed values.  In the small scale  dynamo studied, we use the  turbulent spectrum given by   \citet{sub99}.  The characteristic velocity $V_{c}$ and length $L_{c}$,  used in the expression for the  vorticity $V_{c}/L_{c}$,  are  $V_{c} \simeq 10^{7} cm/s$ and $L_{c} \simeq$ 200 kpc. This vorticity is comparable to  that found by \citet{ryu08},   studying   the formation of large scale structures. The length $L_{c} \simeq 200$ kpc used is a characteristic size of a protogalactic cloud.   The turbulent spectrum used simulates Kolmogorov turbulence \citep{vai82}.  From  our Figs. 1 and 2,  we find that $M_{L}(\sim B^{2})$ increases from $\sim 10^{-23} ~G^{2}$ (corresponding to a magnetic field $B\sim 3\times 10^{-12} ~G$  over a region $L\sim $ 2 kpc) to $M_{L} \sim 10^{18}~G^{2}$ (corresponding to a field $\sim 10^{-9}~G$ over a region $L\sim 2$ kpc)  in $10^{9}$ years. This corresponds to  a $\sim 6$ e-fold amplification of $B$ in a relatively short time. 
Collapsing to form galaxies at redshift $\emph{z}\sim10$, the density increases by a factor of  $\sim 200$ and the   magnetic fields  are amplified by a factor of $\sim 34$. This predicts $0.03~ \mu G$ fields over 0.34 kpc regions  in galaxies.  If the dipole magnetic fields predicted by the Fluctuation-Dissipation Theorem are  not completely random, microgauss fields over regions $> 0.34$ kpc are easily obtained. The model studied is thus a strong candidate  to explain the $\mu G$ fields observed in high redshift galaxies.

\begin{figure}
\centering
\includegraphics[scale=0.55]{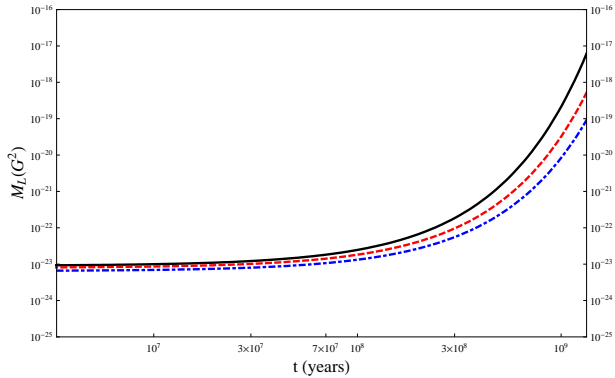}
\caption{Values of $M_{L}(G^{2}$) as a function of t (years) and r. Solid black line has the  reference values: $M_{L}(r,0)= 10^{-11} (0.1pc/r)^{3}~G^{2}$,  $L_{c} = 200 ~kpc$, r = 3 kpc,   and $V_{c}=10^{7}~ cm/s$ in Eqs. (36)-(38).   Dashed red line is for    $r= 4 ~kpc$, and the dotted blue line  for  $r = ~5 kpc$.}
\label{f1}
\end{figure}
\begin{figure}
\centering
\includegraphics[scale=0.55]{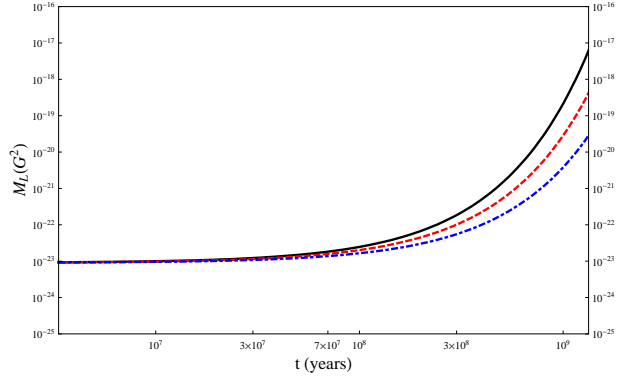}
\caption{Values of $M_{L}(G^{2}$) as a function of t (years),  varying $V_{c}$. Solid black line has the  reference values in  Fig. \ref{f1}.    Dashed red line  is for  $V_{c}= 8\times10^{6}~cm/s$. Dotted blue line is for $V_{c}= 6\times10^{6}~ cm/s$.}
\label{f3}
\end{figure}
\begin{figure}
\centering
\includegraphics[scale=0.55]{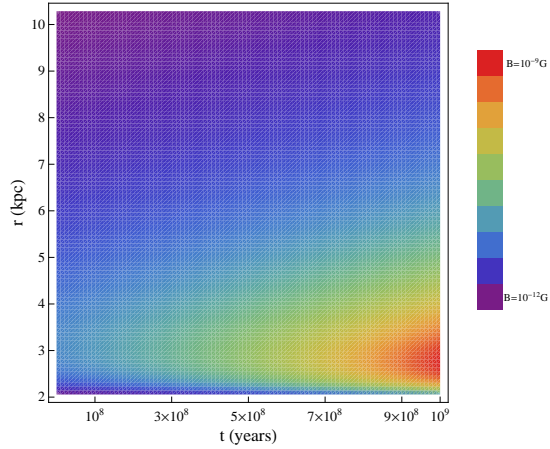}
\caption{Values  of B(G) as a function of t (years) and r(kpc) for reference values of Fig. 1.}
\label{f33}
\end{figure}
\acknowledgments
R.S.S.  thanks the Brazilian agency FAPESP for financial support  (2009/05176-4). R.O.  thanks  FAPESP (00/06770-2) and the Brazilian agency CNPq (300414/82-0) for partial support. We thanks Rainer Beck and Tigran Arshakian for various suggestions. We would also like to thank Joshua Frieman and Wayne Hu for helpful comments.  Finally, we also thank the suggestions of anonymous referee.

\end{document}